\documentclass[iop,numberedappendix]{emulateapj}
\usepackage{natbib}
\usepackage{amsmath}
\usepackage{booktabs}
\usepackage{longtable}
\usepackage{color}
\usepackage{graphicx, subfigure}
\usepackage{hyperref}
\usepackage{ulem}
\begin{document}
\title{Ice mantles on dust grains: dramatic variation of thickness with grain size}
\shorttitle{variation of icy mantle thickness with grain size}
\author{Kedron Silsbee\altaffilmark{1}, Paola Caselli \altaffilmark{1}, \& Alexei V. Ivlev \altaffilmark{1}} 
\altaffiltext{1}{Max-Planck Institute for Extraterrestrial Physics,
Garching by Munich, 85748, Germany; 
ksilsbee@mpe.mpg.de}

\begin{abstract}
We compute the desorption rate of icy mantles on dust grains as a function of the size and composition of both the grain and the mantle.  We combine existing models of cosmic ray (CR) related desorption phenomena with a model of CR transport to accurately calculate the desorption rates in dark regions of molecular clouds.  We show that different desorption mechanisms dominate for grains of different sizes, and in different regions of the cloud.  We then use these calculations to investigate a simple model of the growth of mantles, given a distribution of grain sizes.  We find that modest variations of the desorption rate with grain size lead to a strong dependence of mantle thickness on grain size.  Furthermore, we show that freeze-out is almost complete in the absence of an external UV field, even when photodesorption from CR produced UV is taken into consideration.  Even at gas densities of $10^4$ ${\rm cm^{-3}}$, less than 30\% of the CO remains in the gas phase after $3\times 10^5$ years for standard values of the CR ionization rate.
\end{abstract}
\section{Introduction} 
\label{sect:Introduction}

Interstellar dust grains in molecular clouds provide catalytic surfaces where important molecules such as water and other highly hydrogenated species \citep[e.g. NH$_3$, CH$_{3}$OH, H$_{2}$S;][]{Hasegawa92} can efficiently form. Moreover, dust grains can grow thick icy mantles, where abundant gas-phase species such as CO can freeze out \citep[e.g.][]{Bergin95, Caselli99}. In pre-stellar cores, the precursors of stellar systems, state-of-the-art gas-grain chemical models predict thick icy mantles of a few hundred monolayers around 0.1 $\mu m$ dust particles \citep[e.g.][]{Vasyunin17}. These icy mantles are predicted to contain mainly water and CO, together with CO$_2$, NH$_3$, CH$_3$OH, as observed in molecular clouds and star forming regions \citep[][and references therein]{Boogert15}, as well as organic species of various complexity which will soon be possible to detect with the {\it{James Webb Space Telescope}}  (JWST). Part of this volatile-rich dust will then be incorporated into the protoplanetary disk, which will form during the contraction of dense cloud cores \citep[e.g.][]{Zhao18a}, thus contributing to the build-up of the future planetary system \citep[e.g.][]{Cridland20}. It is therefore crucial to study the main physical mechanisms present in molecular clouds, which could affect the structure and composition of icy mantles at different evolutionary stages.  
\par
Among those mechanisms, particularly important are processes related to cosmic rays (CRs), as they are the only external energy source that can propagate in dense interstellar material almost unhampered \citep{Chabot16, Ivlev18, Padovani18, Silsbee18}. In fact, it has been shown theoretically that direct strikes of low-energy proton CRs on dust grains can produce significant changes in the chemical composition of icy mantles and related gas-phase chemistry \citep{Shingledecker18, Shingledecker20}. In case of collisions with higher energy heavy nuclei, icy mantles could be explosively disrupted, as long as enough radicals are contained within the ice \citep[e.g.][]{Shen04, Rawlings13, Ivlev15}.  Even if mantle explosions do not occur, heavy nuclei can still heat the whole grain to temperatures large enough to allow volatiles to desorb thermally \citep{Leger85, Hasegawa93}, thus producing changes in the thickness and composition of icy mantles. Sputtering by CRs may also be an important desorption mechanism \citep{Leger85, Shen04, Ivlev15, Dartois21}.  Finally, CRs are also responsible for the tenuous UV field produced by the fluorescence of H$_2$ molecules excited by CR-impacts \citep{Prasad83, Gredel89}. These internally-generated UV photons are responsible for the photodesorption of solid species in the top few monolayers of icy mantles \citep[e.g.][]{Shen04, Oberg07, Hollenbach09, Keto10, Cruzdiaz14a, Cruzdiaz14b, Fillion2014, Bertin2016, Fillion21} within dense clouds, allowing species such as water vapor to reach observable abundances even in the central regions of pre-stellar cores \citep{Caselli12b}, where the freeze-out time scale is less than 1000 yr. 
\par
In current gas-grain chemical codes, the interaction of dust grains with CRs is treated in an approximate way, to allow the complex chemistry to unfold. Sputtering has been recently included in the Nautilus gas-grain model by \citet{Wakelam2021}, who conclude that this desorption mechanism can be important in cold gas at volume densities above a few $\times 10^4$\,cm$^{-3}$, but more work is needed to assess the efficiency. The whole-grain heating is usually simplified by assuming a fixed energy deposition within a dust grain with fixed size of 0.1 $\mu$m; another assumption is that much of the volatile desorption occurs near 70\,K \citep[][hereafter HH93]{Hasegawa93}. This assumption is typically adopted by available gas-grain chemical codes \citep[e.g.][]{Garrod_WidicusWeaver2013,Ruaud2016,Sipila2018}, and only recently, advanced treatments have started to be implemented \citep{Sipila2021}. Based on the HH93 suggested formula, \citet{zhao18b} considered the whole-grain heating effect in the case of a more realistic grain size distribution. They found that the variation of the desorption rate for grains with different sizes affects the non-ideal MHD diffusivities (important for the evolution of collapsing magnetized pre-stellar cores); it also produces an interesting differentiation between the freeze-out of CO and N$_2$ molecules, which could explain dense core observations, where N-bearing molecules appear more abundant than CO and CO-related species \citep[e.g.][]{Bergin02, Caselli99, Crapsi07, Redaelli19}. Using a comprehensive gas-grain chemical code and following \citet{zhao18b} for the inclusion of a grain size distribution, \citet{Sipila20} found strong variations in the composition of icy mantles depending on the grain size.
\par
Considering the importance of CRs for the structure and composition of icy mantles and related gas phase chemistry, we decided to study in more detail the whole-grain heating, taking into account a model of cosmic ray transport and relaxing the HH93 assumptions. We also reconsider the sputtering and compare these desorption mechanisms to the CR-induced UV photodesorption. Here, we focus on simple icy mantles (pure CO and pure H$_2$O ice) to keep the formulation simple, but the treatment can be extended to other ice mixtures. 
\par
In Section \ref{sect:desorptionMechanisms}, we provide a detailed description of our method for calculating CR desorption rates.  In Section \ref{sect:DRates}, we combine these rate calculations with the calculations of the local CR spectrum to show the contribution of the different mechanisms to the overall desorption rate.   We perform calculations for different environments, dust size distributions, and dust compositions.  In Section \ref{sect:mantleThickness} we explore the time evolution of the mantle thickness distribution.  In Section \ref{sect:discussion} we discuss some implications of these results.  We present our conclusions in Section \ref{sect:conclusions}.
\section{Desorption mechanisms}
\label{sect:desorptionMechanisms}
We consider three desorption mechanisms: thermal desorption due to whole-grain heating, sputtering, and photodesorption from CR-produced UV photons.  In this work, we use the word ``grain" to denote the inner part of the solid body, which is assumed spherical and may be made of either silicates or carbonaceous material.  Grains are coated in mantles of either CO, or ${\rm H_2O}$.  Let $a$ denote the radius of the grain (without mantle), and $a_{\rm tot}$ be the radius of the grain plus the mantle thickness.
\subsection{Thermal Desorption (Whole-Grain Heating)}
\label{sect:wholeGrainHeating}
When a CR particle passes through a grain, it leaves a narrow hot cylindrical track.  The radius of this track was often taken to be 5 nm \citep[e.g.][]{Leger85}, but has recently been shown to vary significantly (between 1 and 7 nm for protons), depending on the particle energy \citep{ShingledeckerGeant20}.  The deposited energy depends on the CR species, the kinetic energy $E$ of the CR particle, and its impact parameter $b$ (defined as the minimum distance between the straight CR track and the grain center).  For sub-micron sized grains, the time for the grain to lose this heat (either due to modified black-body radiation or to sublimation of mantle material over the entire grain) is long compared to the time over which the heat diffuses through the grain.  Because of this, for this calculation, we assume the grain temperature to be uniform.  In order to determine the desorption rate, we need to determine what fraction of the deposited energy after a given collision is used to sublimate molecules from the grain surface (rather than being radiated away).
\par
 The radiative cooling rate $\dot E_{\rm rad}(a_{\rm tot}, T)$, is given as a function of $a_{\rm tot}$ and temperature $T$ by
 \begin{equation}
 \dot E_{\rm rad}(a_{\rm tot}, T) = 4\pi a_{\rm tot}^3 q_{\rm abs} \sigma T^6,
 \label{eq:dotEthDef}
 \end{equation}
 where $q_{\rm abs}$ is a prefactor for the Planck-averaged absorption efficiency, equal to $0.13$ K$^{-2}$ cm$^{-1}$ for silicate grains, and $0.08$ K$^{-2}$ cm$^{-1}$ for carbonaceous grains \citep[][ Section 24.1.3]{Draine11}.  Presumably the presence of the icy mantles would somewhat alter the value of $q_{\rm abs}$, but as discussed below, the desorption rate does not depend very strongly on the degree of radiative cooling.  
 \par
 Assuming the residence time of a molecule on the surface of a grain to be given by Equation 4.30 of \citet{Tielens10}, and assuming each sublimated molecule to carry away energy $E_b$, we find that the sublimation cooling rate for a mantle dominated by a particular species is given by
 \begin{equation}
 \dot E_{\rm subl}(a_{\rm tot}, T) = 4\pi a_{\rm tot}^2N_s E_b \nu e^{-\frac{E_b}{k_BT}},
 \label{eq:dotEEv}
 \end{equation}
where $N_s$ is the number density of surface sites.  For CO, the binding energy $E_b^{\rm CO} = 858 \, k_B{ \rm K}$ and the rate $\nu_{\rm CO}$ is approximately $6 \times 10^{11}\, {\rm s}^{-1}$ \citep{Acharyya07}.  For H$_2$O, we take $E_b^{\rm H_2O} = 5773$ $k_B$K and a rate $\nu_{\rm H_2O} = 10^{15} \, {\rm s}^{-1}$ from \citet{Fraser01}.
The strong dependence of $\dot E_{\rm subl}$ on $T$ means that the transition between sublimation cooling and radiative cooling occurs over a range of only a few degrees, and the transition temperature (approximately 24 K for CO) is almost independent of $a_{\rm tot}$ and $q_{\rm abs}$.  This also means that modest inaccuracies in calculating the sublimation or radiative cooling rates have little effect on the resulting desorption rates.
\par
We write the number of molecules of mantle lost as a result of whole-grain heating to temperature $T$ as 
\begin{equation}
\Delta N = \frac{V_{\rm tot}}{E_b} \int_{T_{\rm eq}}^T \frac{\dot E_{\rm subl}(a_{\rm tot}, T')\bar C(T')dT'}{\dot E_{\rm rad}(a_{\rm tot}, T') + \dot E_{\rm subl}(a_{\rm tot}, T')},
\end{equation}
where $T_{\rm eq}$ is the equilibrium grain temperature in the absence of transient heating, $V_{\rm tot} = V_{\rm gr} + V_{\rm man}$ is the total (grain + mantle) volume, and $\bar C(T)$ is the mean heat capacity per unit volume, averaged over the mantle and grain material:
\begin{equation}
\bar C = \frac{C_{\rm gr} V_{\rm gr} + C_{\rm man} V_{\rm man}}{V_{\rm tot}}.
\end{equation}

  For the heat capacity of the carbon and silicate grains, we use Equations 9 and 11 from \citet{Draine01}; we note that the prefactor in their Equation 10 should be $n$, rather than $n^{-1}$ (Draine, private communication).  For the heat capacity of the CO and H$_2$O mantles, we use the experimental data from \citet{Clayton32} and \citet{Giauque36}.  While the experimental data for CO only go up to a temperature of 66 K, sublimation cooling is already more than a factor of $10^6$ more efficient than radiative cooling at this temperature (for the grains we consider).  Hence, the heat capacity at higher temperature does not affect the desorption rate.
\par
  Let us introduce the number density of mantle molecules $n_{\rm man} = \rho_{\rm man}/m_{\rm man}$, where $\rho_{\rm man}$ is the density of the mantle material, and $m_{\rm man}$ is the mass of the mantle molecule.  We then obtain the mantle volume loss $\Delta V_i \equiv \Delta N_i/n_{\rm man}$ induced by a CR particle of species $i$, as a function of the CR energy $E$ and impact parameter $b$:
 \begin{equation}
\Delta V_i =  \frac{V_{\rm tot}}{n_{\rm man} E_b} \int_{T_{\rm eq}}^{T_i(E, b)} \frac{\dot E_{\rm subl}(a_{\rm tot}, T')\bar C(T')dT'}{\dot E_{\rm rad}(a_{\rm tot}, T') + \dot E_{\rm subl}(a_{\rm tot}, T')}.
\label{eq:FracVolLoss}
\end{equation}
The maximum temperature $T_i(E, b)$ is determined by the relation 
\begin{equation}
E_{{\rm dep}, i}(E, b) = V_{\rm tot}\int_{T_{\rm eq}}^{T} \bar C(T') dT',
\end{equation}
where 
\begin{equation}
E_{{\rm dep}, i} = 2 Q_{{\rm man}, i} \sqrt{a_{\rm tot}^2 - b^2} + 2(Q_{{\rm gr}, i} - Q_{{\rm man}, i}) \sqrt{a^2 - b^2},
\label{Edepeq}
\end{equation}
with the radicals taken as zero if their arguments are negative.  $E_{{\rm dep}, i}(E, b)$ is the heat deposited by a CR species $i$ with the stopping power $Q_i(E)$, which hits the grain with impact parameter $b$.
\par
In order to calculate $E_{{\rm dep}, i}$, we used the SRIM code \citep{Ziegler10}.  We considered grains made of SiO$_2$ and graphite, with densities 3.5 g cm$^{-3}$ and 2.26 g cm$^{-3}$ respectively.  We considered mantles of CO and $\rm H_2O$ with densities 1 g cm$^{-3}$ and 0.9 g cm$^{-3}$ respectively.   The SRIM code gives us the stopping power $Q_i$ in each material, defined as the energy lost per unit distance travelled.  We multiply $Q_i$ by 0.6, to account for the energy lost to secondary electrons which escape the grain without depositing all of their energy.  This factor of 0.6 should in principle be a function of grain size, with larger grains retaining a higher fraction of the secondaries, however the results in Table B1 of \citet{Leger85} (hereafter L85) show a very weak dependence on grain size.  
\par  In the case of thin mantles and high values of $E_{\rm dep}$ it may be that the equations above suggest $\Delta V > V_{\rm man}$, i.e. the grain loses more mantle material than it has.  In this case, we set $\Delta V = V_{\rm man}$. We verify in Section \ref{sect:continuityApplicability} that in our standard model, only a small fraction of the mantle volume is lost in events which remove the entirety of the mantle.  For the most part, mantle loss may be thought of as a continuous process.  To get the total rate of mantle loss from whole-grain heating, we sum over the different CR species and integrate over the local spectrum $j_i(E)$ for each species, and over the impact parameter:
\begin{equation}
\dot V = \sum_i  \int_{0}^\infty 4\pi j_i(E) dE \int_0^a  \Delta V_i(E, b) 2\pi b db.
\end{equation}
\subsection{Sputtering}
\label{sect:spotHeating}
Even if the CR passage does not deposit sufficient energy to induce sublimation of the mantle from whole-grain heating as discussed in the previous section, there may be some loss of material resulting from processes which occur in close proximity to the grain track.  L85 provide an estimate of the ``spot" desorption rate of CO using a thermal spike model.  However, the treatment presented there overpredicts the desorption measured in experiments for the following reasons.  
\par
First, the vapor pressure assumed by L85 is much too high at elevated temperatures.  Compared with experimental values \citep{Michels52}, Equation 4 in L85 overpredicts the vapor pressure of CO by a factor of 3.9 at 93 K, and a factor of 9.5 at 133 K.  This is because the assumption underlying the exponential dependence of vapor pressure on temperature --- that the density of the material in the gas phase at the equilibrium vapor pressure is small compared to the density in the solid phase --- is violated for temperatures above about 130 K given the vapor pressure curve in L85.  Using L85's expression for the heat capacity, the temperature in the center of the grain track after passage of a 200 MeV iron nucleus (L85's example case) is 275 K, so well beyond the region of applicability of the vapor pressure curve assumed in L85. 
\par
Furthermore, as pointed out in \citet{Bringa99}, near the center of the heated region, the heat transport is not purely diffusive. Coulomb effects may also influence the desorption of material \citep{Bringa02}.  Even given the physical assumptions in the L85 model, the calculation of the resulting sputtering presented there is rather approximate.

\par
In light of these uncertainties, we chose instead to use experimental results.  Following the results in \citet{Dartois21}, for a given projectile and mantle species we use the relation
\begin{equation}
\Delta N = N_0  Q_9^2 \left[1 - \exp{(-N_{\rm lay}/N_d)}\right].
\label{eq:spotDesorption}
\end{equation}
Here $N_{\rm lay}$ is the number of layers of mantle material, and $N_d$ is a characteristic depth of the desorbed material, which also depends on the stopping power $Q_9$ (in units of $10^9 \, {\rm eV \, cm^{-1}}$).  For CO, we use $N_0^{\rm CO}$ = 169 and $N_d^{\rm CO} = 8.9 Q_9^{0.95}$ \citep{Dartois21}.  For H$_2$O we use $N_0^{\rm H_2O} = 13.2$ \citep{Dartois21} and $N_d^{\rm H_2O} = 0.4 Q_9$ \citep{Dartois18}.  In calculating values of $N_0$, we have multiplied the experimentally determined desorption yields by 2 to account for the fact that the CR track intersects the grain surface at two points.  In principle there should be some dependence of the desorption yield on the impact parameter (with grazing impacts desorbing more material) but we have not included this in our analysis.  The desorption rate from sputtering is then given as
\begin{equation}
\dot V = \pi a_{\rm tot}^2 \sum_i \int_0^\infty \Delta V_i(E) 4\pi j_i(E) dE,
\end{equation}
where $\Delta V_i(E) = \Delta N_i(E)/n_{\rm man}$.

 \subsection{UV Desorption}
 \label{sect:UVDesorption}
 CRs travelling through the cloud interact with the ${\rm H}_2$ gas to produce UV photons \citep{Prasad83}.  The production rate of photons in the range of wavelengths $[\lambda, \lambda + d\lambda]$ is equal to $\zeta \chi_{\rm UV}(\lambda) d\lambda$, where $\zeta$ is the local ionization rate of H$_2$.  According to \citet{Prasad83}, $\int_0^\infty \chi_{\rm UV}(\lambda) d\lambda = 0.32$ UV photons per ionization event.  These photons are mostly between 100 and 160 nm \citep{Gredel89}.  In calculating the UV desorption rate, we made the approximation that every photon which is produced is absorbed locally by a dust grain.  The extinction column for UV photons is of order $10^{21}$ cm$^{-2}$, which is small compared with the column densities we are considering, thus justifying our local treatment of the UV field.  Suppose then that we have a distribution of dust grain number density $n_d(a_{\rm tot})$ from $a_{\rm min}$ to $a_{\rm max}$, such that $\int_{a_{\rm min}}^{a_{\rm max}} n_d(a_{\rm tot}) da_{\rm tot}$ gives the total number density of dust grains.  The cross section for a grain of size $a_{\rm tot}$ to absorb UV photons of wavelength $\lambda$ is $C_{\rm abs}(a_{\rm tot}, \lambda)$.  We calculate the absorption cross section for the grains using Mie theory and employ the dielectric functions of the grain material (without consideration of the mantle) from \citet{Draine03}.  The rate at which such photons are absorbed per unit volume by all grains with sizes in the range [$a_{\rm tot}$, $a_{\rm tot}+da_{\rm tot}$] is given by $I(\lambda) n_d(a_{\rm tot}) C_{\rm abs}(a_{\rm tot}, \lambda)da_{\rm tot}$.
To determine the resulting equilibrium intensity $I(\lambda)$ of the radiation field, we note that the number of photons of wavelength $\lambda$ absorbed per unit volume must be equal to $\zeta n_{\rm H_2} \chi_{\rm UV}(\lambda)$.  This allows us to write
 \begin{equation}
 I(\lambda) = \frac{\zeta n_{\rm H_2} \chi_{\rm UV}(\lambda) }{\int_{a_{\rm min}}^{a_{\rm max}} n_d(a_{\rm tot}') C_{\rm abs}(a_{\rm tot}', \lambda) da_{\rm tot}'}.
 \label{eq:UVNormalization}
 \end{equation}
The total rate of photon absorption $\dot n_\gamma(a_{\rm tot})$ by a single grain of size $a_{\rm tot}$ is given by $I(\lambda) C_{\rm abs}(a_{\rm tot}, \lambda)$ integrated over $\lambda$,
\begin{equation}
\dot n_\gamma = \int \frac{\zeta n_{\rm H_2} \chi_{\rm UV}(\lambda)  C_{\rm abs}(a_{\rm tot}, \lambda)d \lambda.}{\int_{a_{\rm min}}^{a_{\rm max}} n_d(a_{\rm tot}') C_{\rm abs}(a_{\rm tot}', \lambda) da_{\rm tot}'}. 
\end{equation}
To turn this into a volume desorption rate, we multiply by the photodesorption yield $Y_{\gamma}$ (defined here as the ratio of desorbed mantle molecules to absorbed photons), and the volume $n_{\rm man}^{-1}$ occupied by a single mantle molecule:
\begin{equation}
\dot V = \frac{Y_\gamma \zeta n_{\rm H_2}}{n_{\rm man}}\int \frac{\chi_{\rm UV}(\lambda)C_{\rm abs}(a_{\rm tot}, \lambda)d\lambda}{\int_{a_{\rm min}}^{a_{\rm max}} n_d(a_{\rm tot}') C_{\rm abs}(a_{\rm tot}', \lambda) da_{\rm tot}'}.
\end{equation}
In our calculations, we assume that $\chi_{\rm UV}(\lambda)$ is uniform between 100 and 160 nanometers, and integrates to 0.32.
\par
  There is a great deal of uncertainty concerning the value of $Y_\gamma$.  \citet{Hollenbach09} cite experiments of \citet{Westley95} in which they find a yield for ${\rm H_2O}$ of between $10^{-3}$ and $8 \times 10^{-3}$.  \citet{Hollenbach09} also suggest that most photodesorption occurs if the photon is absorbed in the first two layers of ice, and the yield is approximately 0.1 times the fraction of photons which are absorbed in the first two layers.  For thick ices, \citet{Oberg07} give $Y_\gamma = 3 \times 10^{-3}$ for CO ice and \citet{Oberg09} give $Y_\gamma = 1.3 \times 10^{-3}$ for H$_2$O ice at low temperature.  This is similar to what was found in \citet{Fillion21}, depending on the exact UV spectrum assumed.  We note that other experimental work has suggested a higher yield for CO.  \citet{Fayolle11} found for CO a yield varying between $6.9 \times 10^{-3}$ and $2.8 \times 10^{-2}$ for photons with energies between 8.2 and 11.2 eV.  \citet{Cruzdiaz14b} find values as high as $5 \times 10^{-2}$ depending on the energy of the incident photon.  These experimental yields are typically defined as the ratio of the flux of desorbed particles to the incident photon flux on a planar surface coated with the molecule in question.  It is not trivial to relate this to the ratio of desorbed mantle molecules to absorbed photons.  
  \par
  Based on the above results for $Y_\gamma$, for CO we use a yield which, for small grains, is equal to 0.1 times the fraction of grain volume which is composed of mantle material and contained in the first two layers.  This yield assumes that photons are equally likely to be absorbed at any location in the grain, and result in the desorption of an average of 0.1 mantle molecules per photon absorbed in the outer two layers. For larger grains, photons are more likely to be absorbed in the outer layer, and thus the yield should decrease, converging to a constant.  Therefore, we then set a minimum yield of $4 \times10^{-3}$.  Although the results from the literature are uncertain, they generally favor the desorption rate for H$_2$O being a factor of a few lower than that for CO.  We therefore assume that the yield for H$_2$O is lower by a factor of three.  Unless the yields were more than an order of magnitude larger than the above values, they would not affect the main results in the present paper.
  
  \par

\section{Desorption Rates}
\label{sect:DRates}
In this section we describe the results of calculations using the theory presented in Section \ref{sect:desorptionMechanisms}.  
\subsection{CR Spectra}
\label{sect:crspectra}
 We consider protons, alpha particles, and the most common isotopes of carbon, oxygen, neon, magnesium, silicon, sulfur, argon, calcium and iron. These heavy nuclei are (with the exception of carbon) included in the study of \citet{Ave08}, which used the balloon-borne TRACER instrument to measure the abundances of cosmic rays at high energies.  We assume carbon to have the same flux as oxygen based on results from the AMS experiment presented in \citet{Jia19}.  We take the abundance of helium from the AMS data \citep{Aguilar15b}.  All these measurements are made at energies of roughly a GeV and above, whereas the CR desorption is dominated by roughly MeV CRs. 
 \par
To estimate the abundances of the various species at lower energies, we assumed the energy spectrum of each CR species outside the cloud to be proportional to that of protons.  We normalized the spectra for each species based on the flux ratio relative to protons at 0.9 GeV.  In all cases where we refer to CRs (energies, loss function, or flux), the energy is measured per nucleon.
   \par
 We use two different CR proton spectra and corresponding rates $\zeta$ for the ionization of molecular hydrogen.  These are the ``low" and ``high" CR spectra from \citet{Padovani18}.  The ``low" spectrum is a fit to the data from the Voyager probe \citep{Cummings16}.  The ``high" spectrum is designed to fit the observations near Earth above 1 GeV, but rises more steeply at low energies to match the ionization rate data from \citet{Indriolo12}.  
 \par
The local spectrum within the cloud is different from the external spectrum because of energy losses within the cloud.  To determine the local spectrum, we used the continuously slowing down approximation discussed in \citet{Padovani18}. We took the loss functions for all species to be proportional to that for protons, but multiplied by $z^2/A$, where $z$ is the charge of the species, and $A$ is the total number of nucleons.  The ionization cross-section was assumed to be proportional to $z^2$.  
\subsection{Standard Grain Population}
\label{sect:standardPopulation}
Our ``standard" grain population consists of SiO$_2$ grains with CO mantles.  We further assume an MRN grain size distribution: $n_d(a) \propto a^{-3.5}$ with 5 nm $\leq a \leq 250$ nm, and that every grain has a constant mantle thickness of 7.6 nm.  This mantle thickness was calculated for a ratio of carbon to hydrogen equal to $1.6 \times 10^{-4}$, \citep{Sofia04}, and assuming all the carbon to be in the form of solid CO. 
\par
The size distribution is relevant because it affects the dust opacity, and therefore the local CR-induced UV field.  Because we are assuming that local UV production is balanced by absorption in dust, the UV field is inversely proportional to the local dust opacity at the relevant wavelength (see Equation \eqref{eq:UVNormalization}).  Hence, if the dust were to have grown significantly, then the UV field would be correspondingly stronger.
\subsection{Environments}
We consider two characteristic regions of a molecular cloud core.  The ``outer" region has a molecular hydrogen number density $n = 10^4$ cm$^{-3}$ and particle column density (counting helium) $N = 10^{22}$ cm$^{-2}$.  The ``inner" region has $n = 10^6$ cm$^{-3}$ and $N = 10^{23.5}$ cm$^{-2}$.  In estimating the visual extinction $A_V$, we use a ratio of 1 magnitude of $A_V$ to $10^{21}$ cm$^{-2}$ of H$_2$ column density.  We note that even in the outer region, assuming the extinction in the UV to be 1.8 times that in the optical, the external UV field is reduced by more than a factor of $10^6$, thus bringing it well below the level of the CR-produced UV field.  We assume a dust temperature $T_d$ of 6 K in the inner region, and 12 K in the outer region, based on models of grain temperature in prestellar cores \citep{Evans01, Zucconi01, Goncalves04, Hocuk17}.  The dust temperature matters because it determines the energy input required to heat the grain to 24 K (at which point significant CO-mantle desorption begins), so a wider range of CRs can result in desorption.  The effect, however is not very significant: the ratio of the energy required to go from 12 K to 24 K to the energy required to go from 6 K to 24 K is 81\%.  We assume a gas temperature $T_g$ (relevant to the deposition rate - see Section \ref{sect:mantleThickness})  of 6 K in the inner region \citep{Crapsi07} and 10 K in the outer region \citep{Keto10}.  

\par
The parameters of our environments are summarized in Table \ref{tab:table1}.  These include the total ionization rates $\zeta_{\rm L}$ and $\zeta_{\rm H}$ of molecular hydrogen, corresponding to the attenuated ``low" and ``high" spectra discussed in Section \ref{sect:crspectra}.
\begin{table}[h!]
  \begin{center}
    \caption{{\bf Environment Parameters}}
    \label{tab:table1}
  \begin{tabular}{c c c }
        - & {\bf Outer Region} & {\bf Inner Region} \\
        \hline
     $n$, cm$^{-3}$ & $10^4$ & $10^6$ \\
     $N$, cm$^{-2}$ & $10^{22}$ & $10^{23.5}$\\
     $A_V$ & 8 & 260 \\
     $T_g$, K & 10 & 6\\
     $T_d$, K & 12 & 6 \\
     $\zeta_{\rm L}$, s$^{-1}$ & $3.4 \times 10^{-17}$ & $2.0 \times 10^{-17}$ \\
     $\zeta_{\rm H}$, s$^{-1}$ & $3.6 \times 10^{-16}$ & $7.4 \times 10^{-17}$ \\
    \end{tabular}
  \end{center}
\vspace{-.5cm}
\end{table}
\subsection{Results}
\begin{figure}[htp]
\centering
\includegraphics[width=1.03\columnwidth]{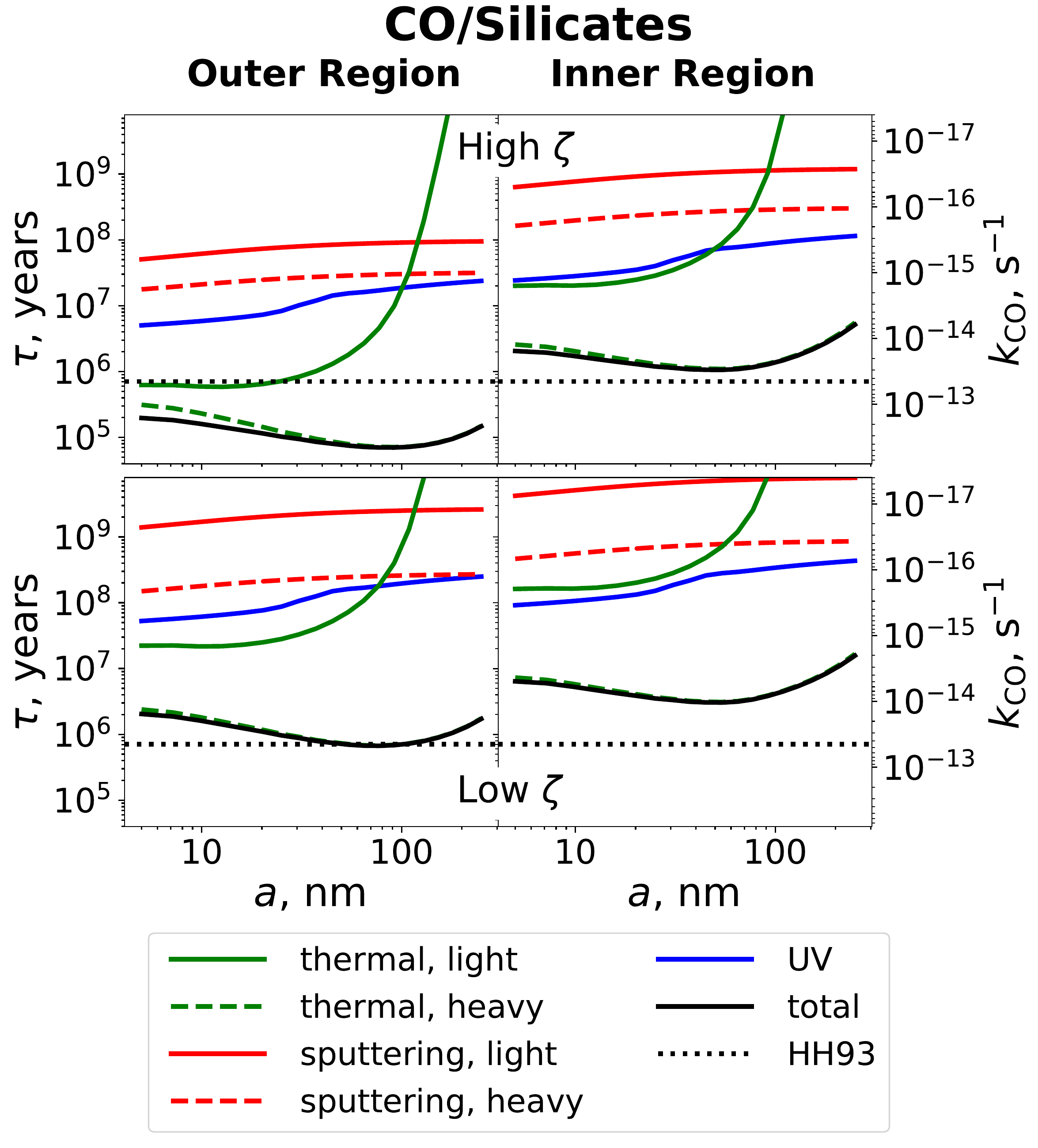}
\caption{Desorption timescale $\tau$ as a function of grain size $a$, assuming an MRN size distribution for the silicate grains and a constant CO-mantle thickness of 7.6 nm.  The rate $k_{\rm CO}$ shown on the right is the inverse of $\tau$.  The solid green line corresponds to thermal desorption (see Section \ref{sect:wholeGrainHeating}) following strikes of light elements (hydrogen and helium), and the dashed green line corresponds to thermal desorption from heavy elements.  The solid and dashed red lines correspond to sputtering from light and heavy elements respectively (see Section \ref{sect:spotHeating}).  The blue line corresponds to photodesorption resulting from the CR-induced UV field in the cloud (see Section \ref{sect:UVDesorption}).  The solid black line shows the total desorption timescale given by Equation \eqref{eq:tauTotal}.  The horizontal dotted line corresponds to the approximation given in HH93 that the grain spends a fraction $\sim 3 \times 10^{-19}$ of their time at 70K, and are otherwise at temperatures too low to desorb any material.}
\label{MRNDesorption}
\end{figure}

 Figure \ref{MRNDesorption} shows the desorption timescales as a function of grain size for our standard grain population (see Section \ref{sect:standardPopulation}).  The legend shows the timescales for desorption due to different mechanisms, as well as the resulting timescale $\tau_{\rm tot}$ for desorption due to all mechanisms discussed in Section \ref{sect:desorptionMechanisms}, given by 
\begin{equation}
\tau_{\rm tot}^{-1}  = \sum_\alpha \tau_\alpha^{-1}.
\label{eq:tauTotal}
\end{equation}
The desorption timescale from mechanism $\alpha$ is
\begin{equation}
\tau_\alpha = V_{\rm man}/\dot V_\alpha,
\end{equation}
where $\dot V_\alpha$ is the corresponding rate of mantle volume loss.  On the right, we show the desorption rate $k_{\rm CO}$ per molecule (i.e. the inverse of the desorption timescale), as would be used in rate equation models such as HH93.
\par
In calculating the desorption rates, we differentiate between ``light CRs" (protons and alpha particles) and ``heavy CRs" (everything else).  Figure \ref{MRNDesorption} shows that desorption is dominated by thermal desorption from heavy CRs.  For small grain sizes, thermal desorption from light CRs also plays a role.  The light elements are not able to heat the larger grains to temperatures high enough to result in significant thermal desorption.  
\par
For these grains, photodesorption is not a dominant process.  This may seem a bit surprising, as of order 10\% of the energy deposited into the cloud by CRs goes into the production of UV photons \citep{Prasad83, Dalgarno99}, all of which, in our model, are absorbed by dust grains.  However, the yield is a few times $10^{-3}$, and also the ratio of desorption energy ($858\,  k_B {\, \rm K} \approx 0.07$ eV) to photon energy $\approx 10$ eV is only around 1\%.  For this reason, we estimate between $10^{-5}$ and $10^{-6}$ of the CR energy which is lost in the cloud goes to desorbing mantle material in this way.  In contrast, about $10^{-3}$ of the CR energy is lost to direct grain heating, and  -- depending on the amount of heat deposited in single events (i.e. whether it is able to heat the grain significantly above 24 K) -- most of it can go to mantle desorption.  
\par
We also show the desorption timescale that arises from the approximation in HH93 as a dotted line.  In this formulation, the grains spend a fraction $3.16 \times 10^{-19}$ of their time at 70 K as a result of CR strikes, and are otherwise assumed to be at temperature low enough that desorption is negligible.  We note however that the binding energy used in this paper for CO is lower than that assumed in HH93, and we assume that desorption can only occur from the surface layer.  The combined effect of these differences it such that our desorption timescale is a factor of a few shorter than that given in Table 4 of HH93.
\par
\begin{figure}[htp]
\centering
\includegraphics[width=1.03\columnwidth]{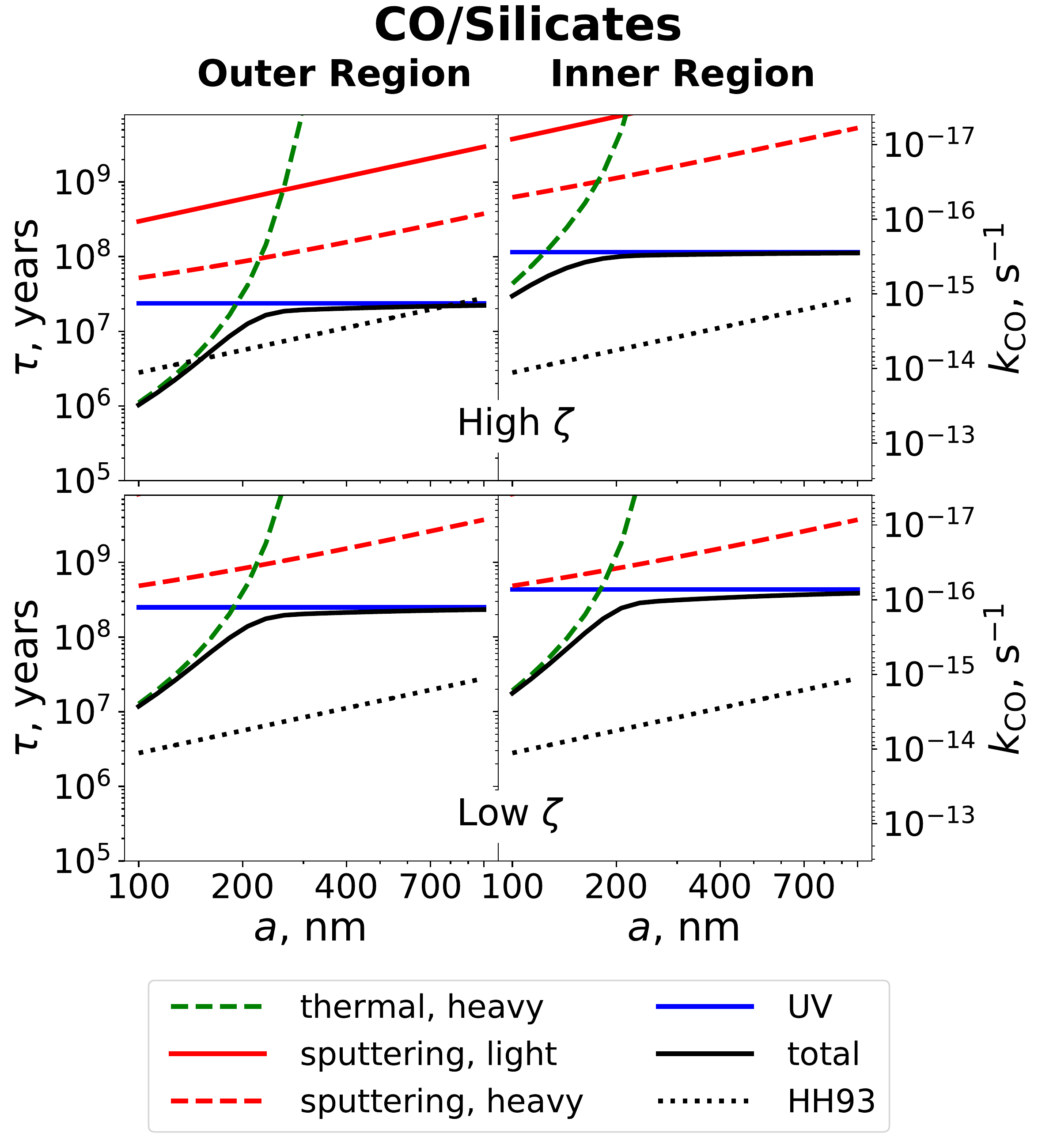}
\caption{Desorption timescales as a function of grain size $a$, assuming a {\it monodisperse} grain size distribution.  The CO-mantle thickness is assumed to be 30\% of the silicate grain radius.  The different color curves have the same meaning as in Figure \ref{MRNDesorption}.  In this case there is no curve for thermal desorption from light particles as the timescale for that process is too long to show on these plots.}
\label{monoDisperseDesorption}
\end{figure}

Figure \ref{monoDisperseDesorption} shows the desorption timescale, again for silicate grains with CO mantles, but this time assuming that there is just one grain size in the distribution.  Coagulation models \citep{Ormel09, Guillet20, Silsbee20} as well as observations \citep{Weingartner01} show that an evolved grain size distribution is typically narrower than the MRN distribution.  In particular, Figures 4 and 5 of \citet{Silsbee20} show that (in the absence of grain fragmentation), grain motions arising from ambipolar diffusion result in dramatic narrowing of the size distribution, particularly in the inner region of the cloud depicted in Figure 4 of that paper.  Here, by using the same mantle-to-grain volume ratio as in the MRN case, we find that the mantle thickness is 30\% of the grain radius.  In this case, for grains larger than 200 nm, CR-induced UV desorption is dominant; however, the desorption timescales in this case are longer than the cloud lifetimes (estimated in \citet{Konyves15} as a few times $10^6$ years for dense cores).  The UV desorption timescale is independent of the grain size in a monodisperse grain size distribution, provided that we assume a size-independent yield.  This is because, irrespective of the grain size, the grains must absorb every UV photon created by CRs.  If the grains grow very large, then the opacity decreases, and the strength of the UV field increases accordingly.
\par
We note that for a given grain size, the desorption timescales are very different between the MRN and monodisperse size distributions.  This is because the heat capacity of CO ice at low temperatures is higher than that of silicate, while the CR stopping power is lower.  For this reason, with the thicker mantles present for the monodisperse size distribution, very few CRs are able to heat grains larger than 100 nm up to the critical temperature where desorption is dominated by sublimation.  Therefore, the thermal desorption timescale increases extremely rapidly with grain size in Figure \ref{monoDisperseDesorption}.  The total desorption timescale becomes determined by the timescale for photodesorption, which is similar in both cases, despite the larger mantles. 

\begin{figure}[htp]
\centering
\includegraphics[width=1.03\columnwidth]{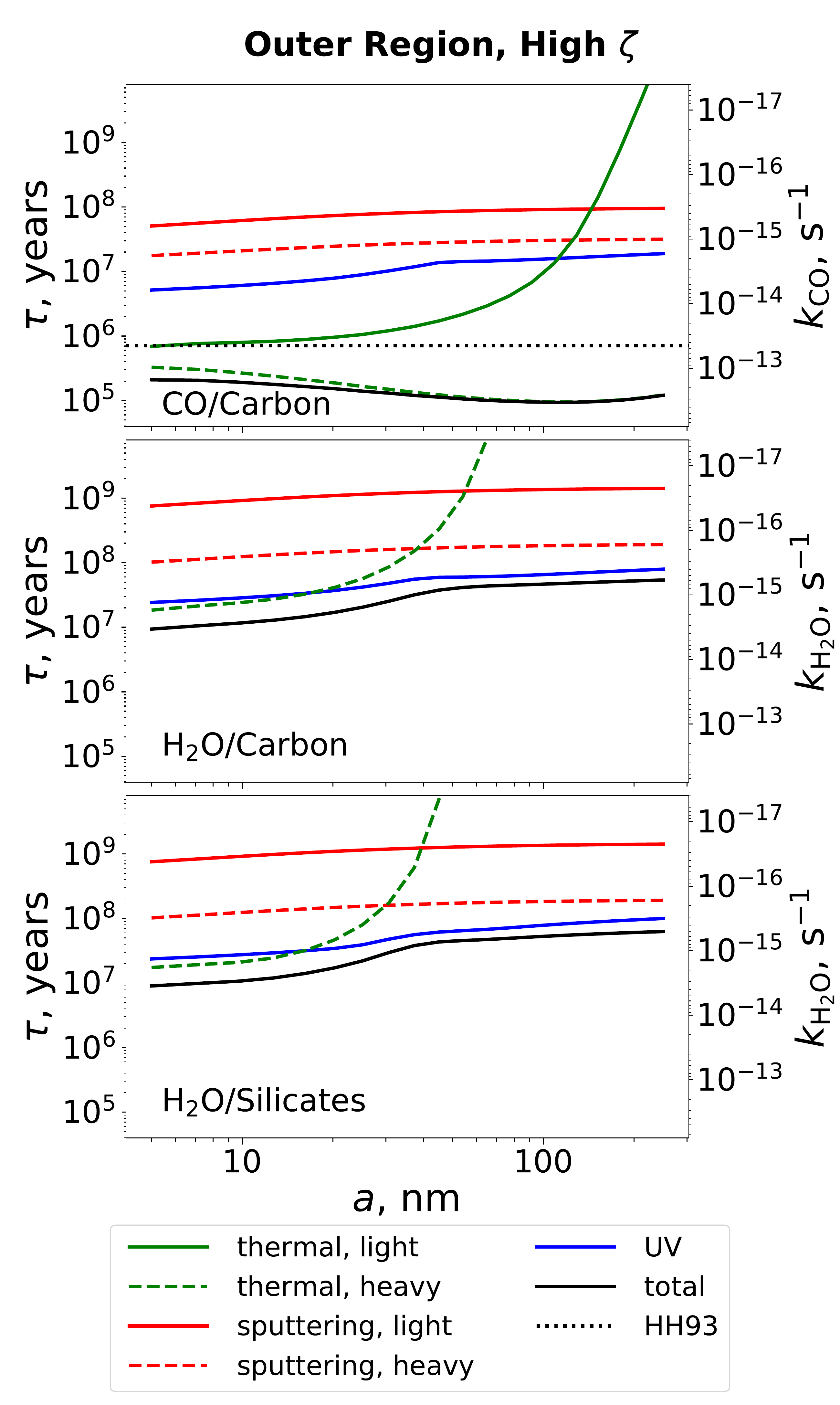}
\caption{Desorption timescales as a function of grain size $a$ for grains of different compositions, as labeled on the panels (e.g., ``CO/Carbon" means a mantle of CO on top of a carbonaceous grain).  As in Figure \ref{MRNDesorption}, we assume an MRN size distribution for the grain and a constant mantle thickness of 7.6 nm.    In all cases we consider the outer region, and the ``high" CR ionization rate.  The different color curves have the same meaning as in Figure \ref{MRNDesorption}.  The curve corresponding to the approximation in HH93 is not plotted for the two panels with H$_2$O mantles because the timescale for desorption of water mantles in that approximation is far too long to show on our plots.}
\label{exoticDesorption}
\end{figure}
\subsubsection{Effect of Grain Composition}
\label{sect:variedComposition}
We also consider carbonaceous grains and water mantles.  The grain composition affects the desorption rate because it determines both how much energy is deposited into the grain, as well as the heat capacity.  The mantle composition is extremely important because the desorption energy of different mantle species varies by factors of several, thus changing the balance between sublimation and radiative cooling.  
\par
We note that icy mantles in molecular clouds contain a mixture of species (mainly H$_2$O, CO, and CO$_2$), while here we present the case of pure ices. This choice allows us to maintain the problem treatable. However, the basic ideas presented here can be applied to more complex chemical networks, which follow the evolution of icy mantle composition in dense clouds. 
\par
Figure \ref{exoticDesorption} shows the desorption timescales for grains of different compositions.  As in Figure \ref{MRNDesorption}, we assume an MRN grain size distribution, and a constant mantle thickness of 7.6 nm.  We consider only the outer-region high-CR model, which corresponds to the top left panel of Figure \ref{MRNDesorption}. 
\par
 The top panel of Figure \ref{exoticDesorption} shows the desorption rates in the case of carbonaceous grains with CO mantles.  This is very similar to the top left panel of Figure \ref{MRNDesorption}, except that the desorption timescales are slightly shorter for larger grains (owing to the reduced heat capacity that more than compensates the reduced energy deposition).  For smaller grains, the heat capacity is dominated by the mantle in either case, and the carbon grains have a slightly slower desorption because of the reduced energy deposition.  These differences are only 7\% and 22\% for grain sizes of 5 and 250 nm respectively.
\par
 The middle and bottom panels of Figure \ref{exoticDesorption} show the timescales for the desorption of ${\rm H_2O}$ mantles from grains composed of carbon and silicates.  We see that the photodesorption from CR-induced UV dominates (even though the yield is rather uncertain), and the timescales are comparable to or longer than the lifetime of a molecular cloud.
\section{evolution of the mantle thickness}
\label{sect:mantleThickness}
We used the desorption rates calculated in Section \ref{sect:DRates} to perform simulations of the growth of mantles. In each case, we begin with an MRN grain size distribution, and assume all the mantle material to be initially in the gas phase.  As before, we assume a sufficient amount of mantle material to be available to form 7.6 nm mantles (if the thickness were the same on all grains).  To calculate the rate of mantle growth, we assumed a geometric cross-section for gas-grain collisions, and that gas phase molecules have a 100\% sticking probability upon contact.  The relative velocities are dominated by the thermal velocity of the gas species.  We pre-tabulated the desorption rates from whole-grain heating and sputtering as a function of grain size and mantle thickness.  These tables are available as supplementary material.
\par
\begin{figure}[htp]
\centering
\includegraphics[width=1.03\columnwidth]{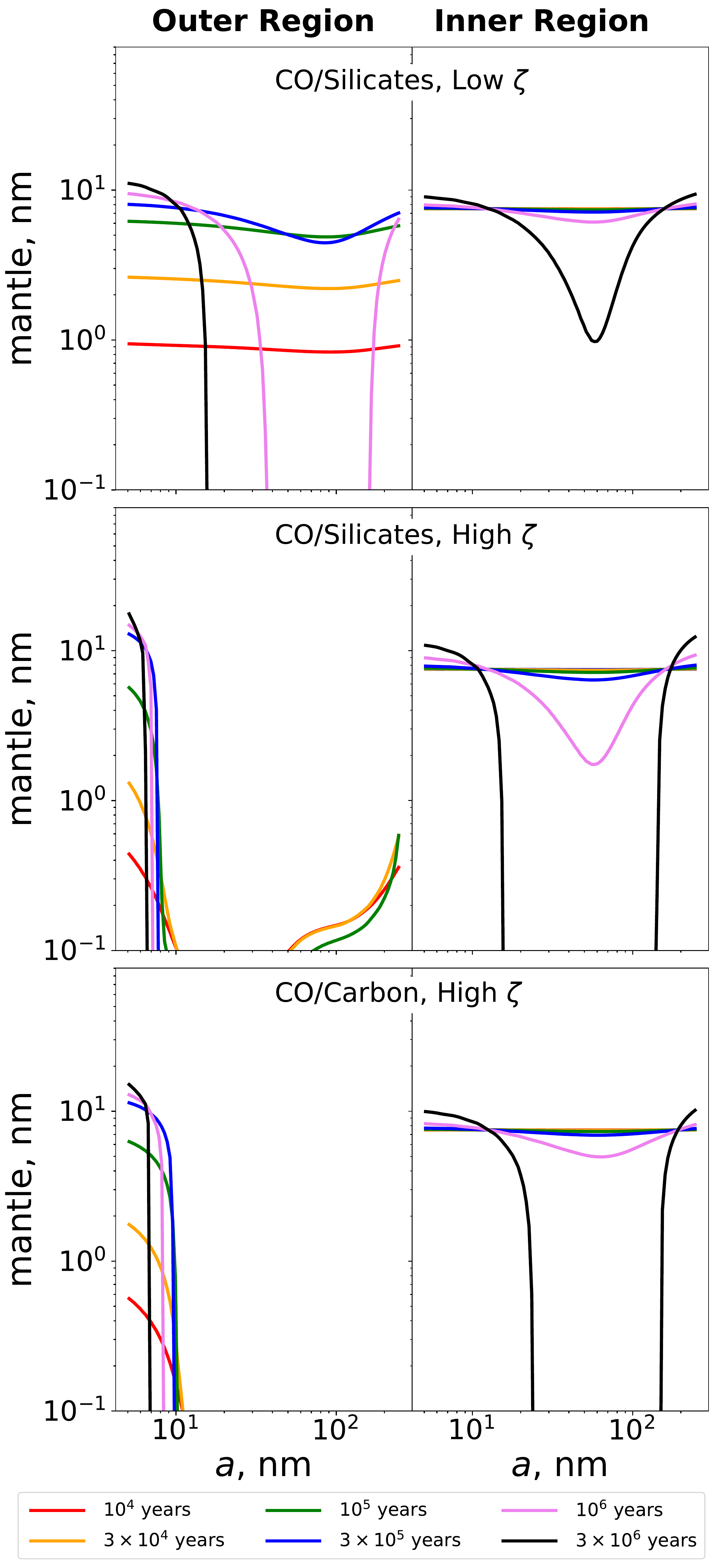}
\caption{Mantle thickness as a function of grain size $a$ assuming an MRN size distribution and that all mantle material is in the gas phase at time $t = 0$.  Mantle growth is regulated by a balance between the sticking of molecules to grains in collisions, and the desorption of molecules from all mechanisms described in Section \ref{sect:desorptionMechanisms}.  The panels on the left correspond to our outer region, and the panels on the right to the inner region.  Different color curves correspond to different amounts of time the simulation was run for, as labelled in the legend.  The composition of the grain and the mantle are labelled on the panels, and the assumed CR ionization rate is shown on the right side of each row.  Because the desorption rate goes to zero as the mantle thickness goes to zero, there is a residual mantle even in the ``gaps", but its thickness is much less than one monolayer.}
\label{megaFigure}
\end{figure}

Figure \ref{megaFigure} shows the mantle thickness as a function of grain radius at various snapshots of time during the simulation.   Each panel corresponds to a different environment, grain composition, and ionization rate, as labelled. 
\par
Remarkably, we find that given enough time, drastic variations in mantle thickness develop, even though the variations in the corresponding desorption timescales are quite modest.  Depending on how the desorption rate varies with mantle thickness, it is possible for grains of some sizes to be almost bare, even though material is only desorbed slightly faster than from grains of other sizes.  
\par

For our typical parameters, the rate at which material is desorbed from the surface is the highest for very thin mantles of about a single layer.  This is the result of a tradeoff between the increased heat capacity which comes from larger mantles (and protects the grain from desorption) and the fact that CRs with higher stopping powers may desorb the entire mantle if it is thin (thus wasting some of the energy).  The fact that the volume loss rate peaks for very thin mantles means there is a runaway effect, and minor variations in mantle thickness become amplified.
\par
\begin{figure}[htp]
\centering
\includegraphics[width=1.03\columnwidth]{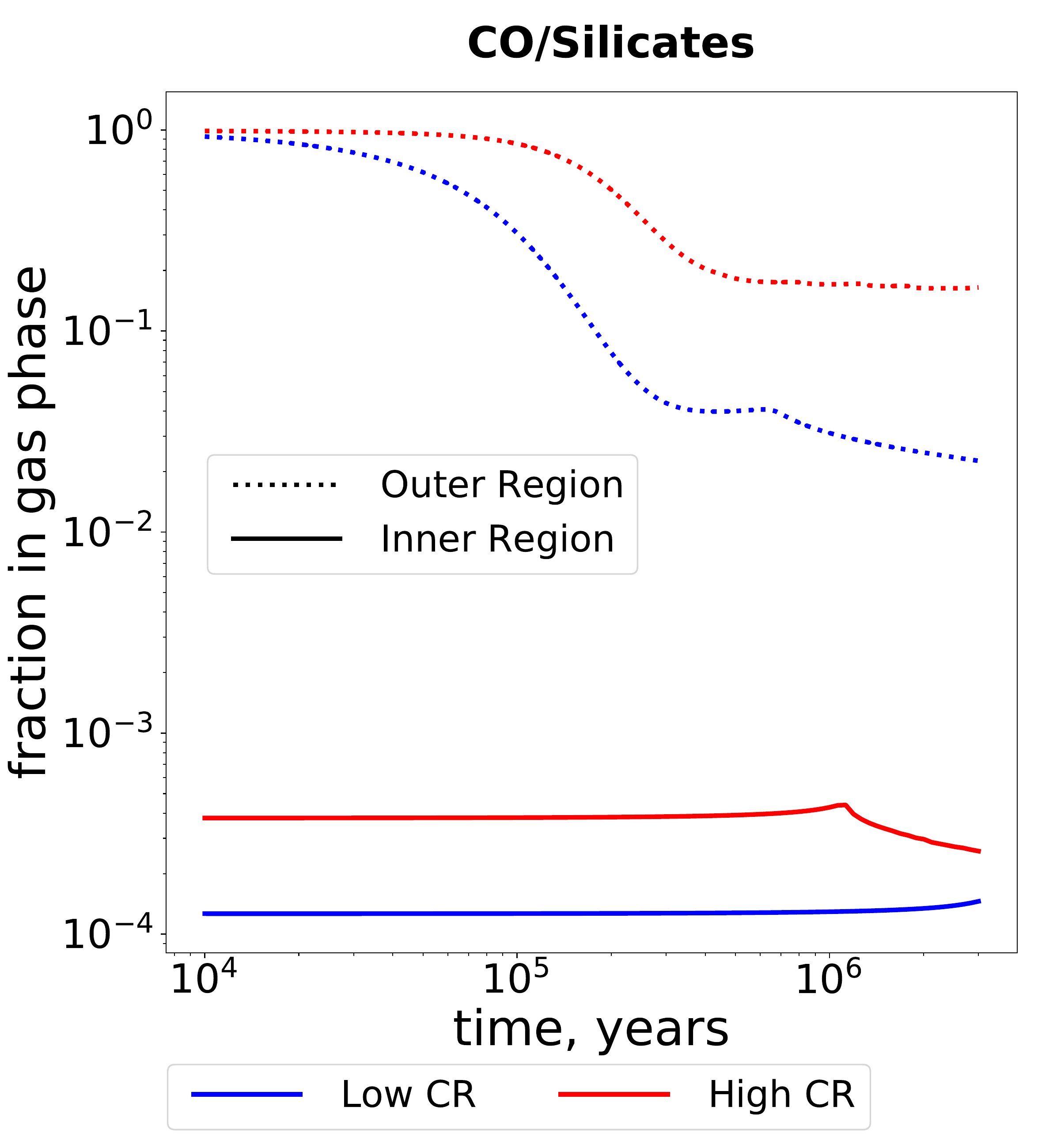}
\caption{Fraction of the CO remaining in the gas phase as a function of time for different simulation runs.  In all cases, we consider CO mantles on silicate dust grains.  The latter have an MRN size distribution.  Different curves correspond to different CR models and regions of the cloud, as labelled in the legend.  In all models considered, over 70\% of the  CO is deposited on the grains by $3 \times 10^5$ years.  The kinks visible in some of the curves around $10^6$ years correspond to the time at which the mantle is lost from grains of intermediate sizes, as shown in Figure 4.   }
\label{gasAbundances}
\end{figure}

Figure \ref{gasAbundances} shows the fraction of the CO which is in the gas phase for each of the simulations with silicate grains and CO mantles, shown in the top and middle rows of Figure \ref{megaFigure}.  We note that freeze-out in these models is almost complete given enough time, consistent with the observations in \citet{Caselli99}.  Using the ``high" model for $\zeta$, only 28\% of the CO is in the gas phase after $3 \times 10^5$ years at a density of $10^4$ cm$^{-3}$.  We note that the evolving mantle size distribution plays a crucial role in the evolution of the gas-phase abundance.  For example, with the high $\zeta$ at $n = 10^4$ cm$^{-3}$, the depletion is slow at the beginning, but accelerates as the growth of mantles on the smallest grains lowers their desorption rate.  We conclude that CR desorption cannot prevent the efficient freeze-out of CO at densities of $10^4$ cm$^{-3}$ or above.  The processes regulating the density at which freeze-out occurs are discussed in Section \ref{sect:freezeOut}.

\subsection{Effect of Grain Composition}

\par
We do not plot the growth of ${\rm H_2O}$ mantles because the desorption timescales are so slow that nothing interesting happens within $3 \times 10^6$ years.  Essentially all the ${\rm H_2O}$ freezes out onto the grains, and then the mantle thickness remains unchanged.
\par
The bottom row of Figure \ref{megaFigure} shows the growth of CO mantles onto carbonaceous grains.  Comparing with the middle row, we see a similar behavior for small grains, but the mantles do not grow at all for grains above $\approx 10$ nm, as the desorption time is a bit faster for the largest grains (see Section \ref{sect:variedComposition}).

\section{discussion}
\label{sect:discussion}
\subsection{Applicability of Continuous Desorption Approximation}
\label{sect:continuityApplicability}

In this paper we have treated the desorption as though it were a continuous process.  However, if a significant amount of the desorption were occurring due to events which remove a large fraction of the mantle, then the problem would become significantly more complicated, as we would need to consider a distribution of mantle thicknesses at each grain size.  Such ``catastrophic desorption" events are more prominent for smaller grains.  For our standard grain population (MRN size distribution with silicate dust grains and CO mantles of thickness 7.6 nm), we found for all regions and grain sizes considered, less than 10\% of the mantle volume is lost in events which remove at least half of the mantle for the smallest grains.  This allows us to conclude that mantle loss can be modelled as a continuous process.
\subsection{Bulk Diffusion Due to Transient Whole-Grain Heating}
It has been suggested \citep{Garrod13} that there could be a diffusion of molecules through the bulk of the icy mantles.  This has also been included in the chemical modeling done in \citet{Vasyunin17}.  The diffusion is assumed to be the result of a ``swapping" process, in which two neighboring molecules swap positions in the bulk ice.  The rate of this process is described \citep{Garrod13} by a rate equation 
\begin{equation}
\tau_{\rm swap}^{-1} = \sqrt{\frac{2N_s E_b}{\pi^2 m}} e^{\frac{E_{\rm swap}}{k_BT}}.
\end{equation}
The value of $E_{\rm swap}$ is extremely uncertain.  For diffusion occurring on the surface of the ice, the ratio of $E_{\rm swap}$ to the binding energy $E_b$ has been estimated by various authors \citep[see][and references therein]{Garrod11} to be in a range between 0.6 and 1.6 (using the assumption of \citet{Garrod13}, who estimated $E_{\rm swap}$ in the bulk to be twice that on the surface).  We assume it takes approximately $3 N_{\rm lay}^2$ swaps before the position of a molecule is randomized within the ice.  Depending on the value of $E_{\rm swap}$, this process may be either dominant in the sense that $3 N_{\rm lay}^2$ swaps are achieved in a time short compared with the chemical evolution timescale for the cloud, or it can be completely unimportant in the sense that the time to achieve even one swap is greater than the lifetime of the cloud.  For example, for a 0.1 micron grain with a 10 nm CO mantle, if $E_{\rm swap}/E_b = 0.6$, then the time for randomization is just 2700 years in the outer region with the high CR ionization rate.  However, this increases to $2 \times 10^7$ years for $E_{\rm swap}/E_b = 1.0$.  If the CO is replaced with H$_2$O, then the timescales even for one swap are all longer than the age of the Universe, even for $E_{\rm swap}/E_b = 0.6$.  It is worth noting also that \citet{Shingledecker19} compared chemical models with experiments, and found support for a non-diffusive model of bulk chemistry.
\subsection{Possibility of explosive desorption}
It was pointed out in \citet{Shen04}, that the energy absorbed in the ice from CR-generated UV photons is a factor of 10 higher than that absorbed by direct impact of CRs.  They propose a model in which this energy is then explosively released when a CR of sufficient energy strikes the grain \citep[see also][]{Greenberg76, Rawlings13}.  This could in principle be a significant source of additional desorption, particularly for larger grains.  However, as pointed out in \citet{Ivlev15}, there is a great deal of uncertainty regarding whether a high enough density of reactive species actually develops in the ice for this desorption mechanism to operate.  We therefore ignore such desorption in our calculations, though it could in principle be at least as important as the other mechanisms, particularly for larger grains.

\subsection{Onset of Freeze-out}
\label{sect:freezeOut}
As shown in Figure \ref{gasAbundances}, already at a density of $10^4$ cm$^{-3}$ desorption from CR-related processes is insufficient to prevent freeze-out of CO.  Indeed, H$_2$O and CO ice have been found to be present in molecular clouds above visual extinctions of three and six magnitudes, respectively \citep[][Figure 7]{Boogert15}.  It is interesting to ask then, what does regulate the onset of freeze-out.  Two processes not considered in this work are thermal desorption at the equilibrium temperature, and the existence of the external UV field.  To calculate desorption resulting from the equilibrium temperature for CO, we take Equation \eqref{eq:dotEEv}, and divide by $4\pi a_{\rm tot}^2 \Delta E$ to get the number of molecules sublimated per unit time per unit area.  Dividing this by $N_s \sim n_{\rm CO}^{2/3}$ to get the number of layers sublimated per unit time, we find even for our low-density dust temperature of 12 K it would take over $10^{11}$ years to sublimate one layer; however, this timescale reduces to $3.5 \times 10^5$ years at 15 K and $400$ years at 17 K.  Based on the curve of dust temperature vs. visual extinction presented in \citet{Hocuk17}, we can conclude that thermal desorption arising form the equilibrium grain temperature is not significant at visual extinctions greater than 1 magnitude, but this conclusion could be changed by a temperature deviation of only a few degrees. 
\par
The external UV field also plays an important role in regulating the onset of freeze out.  Using a total UV flux of $2 \times 10^7 e^{-1.8{\rm A_V}} G_0 $ photons cm$^{-2}$ s$^{-1}$ sr $^{-1}$  \citep{Draine78, Hollenbach09}, approximating the UV photons to have an energy of 10 eV, and using a yield of 0.004 we obtain a desorption rate of $1.3 \times 10^{-4}$ particles per second from a 0.1 micron grain.  Comparing this to the deposition rate, assuming $3.2 \times 10^{-4}$ gas phase CO molecules per ${\rm H_2}$ molecule, we find that freeze-out is faster than UV desorption when $n_{\rm H_2} e^{1.8 \rm A_V} \geq 10^5 G_0$ cm$^{-3}$.  
\par
At sufficiently low densities, the mantle formation can also be limited by CR thermal desorption.  Indeed, assuming an MRN size distribution, the deposition timescale is approximately $ 10^6$ $\times \left(n_{\rm H_2}/10^3\, {\rm cm}^{-3}\right)^{-1}$ years.  Comparing this with the CO-desorption timescales shown in Figure \ref{MRNDesorption}, it is clear (taking into account that the CR desorption timescale increases with gas density due to CR attenuation) that CRs alone can prevent CO freeze-out at around a density of a few times $10^3$ cm$^{-3}$, depending on the CR model.
\par
We conclude that depending on the CR ionization rate, gas density and strength of the interstellar UV field, either whole-grain heating from CRs or photodesorption from interstellar UV is likely to determine at which location freeze-out begins.  This has profound consequences on the properties of the ice mantles expected in the cloud.  If mantles form under conditions in which desorption is dominated by interstellar UV, then the mantle structure will depend on the relative magnitude of the photodesorption rates, becoming rather homogeneous in the case that those rates are similar.  On the other hand, if desorption is dominated by CRs, then the species with higher binding energies, such as water, will form and stay (or freeze-out) on the surface, forming a first layer which will be later covered by species with lower binding energies.
\section{conclusions}
\label{sect:conclusions}
We used existing results from the literature to perform a careful calculation of the rate of desorption of mantle material from dust grains in dark regions of molecular clouds.  We calculated rates for thermal desorption from whole-grain heating, sputtering, and photodesorption from CR-produced UV photons.  We considered CR protons, alpha particles, and the main heavier nuclei (carbon, oxygen, neon, magnesium, silicon, sulfur, argon, calcium and iron).  We derived the local spectra for each of these species using the continuous slowing-down approximation \citep{Padovani18}.  We explored how the rates varied with grain size, grain composition, gas density, and CR ionization rate.  We used these desorption rates to run numerical simulations of the growth of mantles on a population of grains with different sizes.  The tabulated desorption rates due to whole-grain heating and sputtering for grains with CO mantles as a function of grain size and mantle thickness are available as supplementary material.  Based on these studies, we reach the following conclusions.
\begin{enumerate}
\item{Given a standard model of CR propagation, the CR desorption rate decreases by approximately an order of magnitude between a column density $10^{22}$ cm$^{-2}$ and $3 \times 10^{23}$ cm$^{-2}$. }
\item{Assuming a constant mantle thickness of 7.6 nm, the timescale for desorption of CO as a function of grain size has a minimum around 50 to 100 nm.  This is similar to what is seen in Figure 7 of \citet{Leger85} in the limit of thin mantles.  Smaller grains are hit less frequently by CRs, and are dominated by mantle material (which absorbs less energy from the CRs due to the lower stopping power).  Larger grains desorb less efficiently because many CR impacts do not heat them to sufficiently high temperatures to efficiently desorb their mantles.  Grains larger than about 200-500 nm, depending on the mantle thickness, have negligible thermal desorption rates.}
\item{Small (less than a factor of two) differences in the desorption rate for grains of different sizes can, over million year timescales result in an extremely uneven distribution of mantle thicknesses for more volatile species, such as CO. }
\item{CR-induced desorption alone cannot prevent the efficient freeze-out of CO mantles, even at a density of $10^4$ cm$^{-3}$.  This is consistent with observations of solid CO and H$_2$O in molecular clouds such as Taurus, with average densities of $10^3$ cm$^{-3}$.  However, this does not mean such desorption is irrelevant, as it still mediates the exchange of material between the solid and gaseous phases at much higher densities.  Depending on the strength of the CR and UV fields, the freeze-out at somewhat lower densities may also be regulated by photodesorption due to interstellar UV photons.  The structure of the ice mantles may dramatically depend on which of the two desorption mechanisms dominates.}
\item{Whole-grain heating from CRs does not contribute appreciably to the desorption of $\rm{H_2O}$ from mantles.  Because of the high desorption energy of ${\rm H_2O}$, CRs are unable to heat the grains to sufficiently high temperatures for much $\rm{H_2O}$ to be sublimated.  Therefore, $\rm{H_2O}$ desorption is determined by photodesorption from CR-produced UV photons and sputtering (with UV dominating for the yields we chose), and the timescales are on the order of $10^7-10^8$ years. }
\item{In realistic conditions, grains develop mantles with multiple chemical components.  Models suggest \citep{Kalvans15} that the more volatile materials are distributed in the surface layers.  The results of our calculation show that these more volatile species will be almost completely gone for a subset of the grains.}
\end{enumerate}
\section*{data availability}
We have included a set of 6 tables giving the desorption rate due to whole-grain heating and sputtering as a function of grain and mantle sizes, for different grain compositions and environments.  In addition to these tables are a document explaining how to interpret them, and two lists of grain and mantle sizes.
\bibliographystyle{apj}
\bibliography{apj-jour,desorptionRateComparisons}
\end{document}